 \documentstyle[psfig,aps,preprint,tighten]{revtex}
\begin{document}
\draft
\preprint{\vbox{\it Submitted to Nucl. Phys. A \hfill\rm TRI-PP-95-87}}

\title{Quasifree Eta Photoproduction from Nuclei}
\author{Frank X. Lee}
\address{TRIUMF, 4004 Wesbrook Mall,
Vancouver, British Columbia, Canada V6T 2A3}
\author{Louis E. Wright}
\address{Institute of Nuclear and Particle Physics,
Department of Physics, Ohio University, \\ Athens, Ohio 45701, USA}
\author{Cornelius Bennhold}
\address{Center for Nuclear Studies, Department of Physics,
The George Washington University, \\ Washington, DC 20052, USA}
\author{Lothar Tiator}
\address{Institut f\"ur Kernphysik, Universit\"at Mainz,
55099 Mainz, Germany}

% a different format for author/address list 
%\title{Quasifree Eta Photoproduction from Nuclei}
%\author{Frank X. Lee\footnotemark[1], Louis E. Wright\footnotemark[2],
%Cornelius Bennhold\footnotemark[3], Lothar Tiator\footnotemark[4]}
%\address{\footnotemark[1]TRIUMF, 4004 Wesbrook Mall,
%Vancouver, British Columbia, Canada V6T 2A3 \\
%\footnotemark[2]Institute of Nuclear and Particle Physics,
%Department of Physics, Ohio University, \\ Athens, Ohio 45701, USA \\
%\footnotemark[3]Center for Nuclear Studies, Department of Physics,
%The George Washington University, \\ Washington, DC 20052, USA \\
%\footnotemark[4]Institut f\"ur Kernphysik, Universit\"at Mainz,
%55099 Mainz, Germany}

\date{\today}
\maketitle

\begin{abstract}
Quasifree $\eta$ photoproduction from nuclei is studied in the Distorted
Wave Impulse Approximation (DWIA). The elementary eta production
operator contains Born terms, vector meson and nucleon
resonance contributions and provides an excellent description of the
recent low energy Mainz measurements on the nucleon.
The resonance sector includes the $S_{11}(1535)$,
$P_{11}(1440)$ and $D_{13}(1520)$ states whose couplings are fixed by
independent electromagnetic and hadronic data.
Different models for the $\eta N$ t-matrix are used to construct a
simple
$\eta A$ optical potential based on a $t \rho$-approximation.  We find that
the exclusive $A(\gamma,\eta N)B$ process can be
used to study medium modifications of the $N^*$ resonances, particularly if the photon asymmetry can be measured.
The inclusive $A(\gamma, \eta)X$ reaction is compared
to new data obtained on $^{12}C$, $^{40}Ca$,
and is found to provide a clear distinction between different models for
the $\eta N$ t-matrix.
\end{abstract}

\vspace{1cm}

\parskip=2mm
\section{Introduction}

Photoproduction of the isoscalar eta meson with mass of approximately 550 MeV
offers the
possibility of investigating the $\eta N$- and $\eta A$-interaction, as well
as the basic production amplitude.
Recently it has been suggested by Wilkin\cite{Wil93} that the
very large production
cross sections found in the $pd\rightarrow \eta ^3He$ reaction near threshold
could be due to an $\eta N$ scattering length that is much
larger than the value extracted from  a coupled-channels analysis of the
$\pi^-p\rightarrow\eta n$
 and $\pi N\rightarrow\pi N$ data\cite{Bha85}.  Based on the
K-matrix formalism, Wilkin was able to reproduce the strong threshold
energy dependence of $pd \rightarrow \eta ^3He$ by assuming an $\eta N$
scattering length with a real part more than twice as large as in previous
analyses.  Should this conclusion turn out to be true, it would have
dramatic implications for the $\eta$-nucleus interaction. Most importantly,
a larger value for the scattering length
might indicate a larger probability for the
presence of a ``bound'' $\eta$-state for lighter nuclei such as $^3$He than
had been expected. In fact, this seems to be required to explain the
cusp-like structure seen for near-threshold production in the
$pd\rightarrow {^3He} X$ reaction for a missing mass close to the
$\eta$-mass.
%An uncertainty in Wilkin's analysis arises from a cross
%section measurement of the $p d \rightarrow \eta ^3He$ process very close to
%threshold ($\sim$200 keV) that seems to contradict his description of the
%energy dependence. However, if this one point - which suffers from large
%systematic uncertainties - is ignored, impressive agreement with the
%experiment is achieved. A new experiment at Saclay has already been
%approved to explore the energy dependence in the region very close to
%threshold\cite{Bris94}.
This effect may also remove the discrepancy between experiment
and theory for the reaction $^3He(\pi^-,\eta)^3H$\cite{Kam93}.

Over the last several years there has been renewed interest in the production
of $\eta$--mesons with protons, pions and electrons and their interaction with
nucleons and nuclei. One of the first $\eta$--nuclear experiments, performed at
SATURNE in 1988\cite{Ber88}, reported surprisingly large eta production rates
 near
threshold in the reaction $p(\vec{d},^3He)\eta$. These large cross sections
permitted not only a more precise determination of the $\eta$--mass\cite{Plo92}
but were also used to perform rare decay measurements of the eta\cite{Kes93}.
Additional experiments involving pion induced eta production were performed
at Los Alamos\cite{Pen89}. Again, the experimental cross sections at threshold
region of the reaction $^3$He($\pi^{-},\eta$)$^3$H are above the theoretical
calculations\cite{Kam93,Pen89}.

The advent of high duty-cycle electron accelerators opens for the first time
the opportunity to study the reactions $N(\gamma,\eta)N$ and $N(e,e'\eta)N$
in greater detail. Our present knowledge of the $(\gamma,\eta)$ process is
based solely on some old measurements around 20 years ago\cite{Del67}, along
with very few more recent data from Bates\cite{Dyt90} and Tokyo\cite{Hom88}.
Over the last two years
eta photoproduction from the nucleon has been measured at Mainz and at
Bonn with an accuracy of more than an order of magnitude better than 
in older experiments.
At Mainz, the TAPS collaboration has obtained high-quality data for
angular distributions and total cross sections
for photon energies between threshold and 790 MeV
both on the nucleon~\cite{Kru95} and on nuclear targets 
$^{12}C$, $^{40}Ca$~\cite{Roeb95}.
That may be considered to be
a qualitative breakthrough in the experimental field.
Similarly, data at the higher energies up to 1150 MeV 
will be provided soon by the Phoenics collaboration at ELSA \cite{Br94}.

In this work we present a study of the {\em quasifree} eta photoproduction 
on nuclei $A(\gamma,\eta N)B$ (where $B$ is some discrete nuclear state) 
in the Distorted Wave Impulse Approximation (DWIA). 
The model allows for the study of the production process and the 
final-state interactions without being obscured by the details of 
nuclear transition densities, unlike the reaction of the type 
$A(\gamma,\eta)B$ where nuclear transition densities play an important
role.  This is mainly due to the quasifree nature of the reaction:
the momentum transfer to the residual nucleus can be made small.
The key ingredients of the model are a) the single particle wave
functions and spectroscopic factors, 
b) the elementary eta photoproduction amplitudes, and c) 
the final state interactions. Information on each ingredient can be 
taken directly from independent studies.
Previous studies of such quasifree reactions on the pion have proven 
to be successful~\cite{liwb,cap,eepin}.
Other studies on eta photoproduction from nuclei can be found, for 
example, in  Refs.~\cite{car,ben2,chen,bar,von,hal,hom94}.

In Sec.~\ref{free} we briefly describe the $(\gamma,\eta)$ operator 
and show comparisons with new data from Mainz.
The DWIA formalism for quasifree eta
photoproduction on nuclei is derived in Sec.~\ref{dwia}.
In Sec.~\ref{exclusive} we present predictions for 
differential cross sections of the exclusive reaction 
$A(\gamma,\eta N)B$.
In Sec.~\ref{inclusive} we compare our calculations with new 
data from Mainz on the inclusive reaction $A(\gamma,\eta)X$.
In Sec.~\ref{sum} we summarize our findings and present a brief outlook.
The transformation of CGLN amplitudes from the c.m. frame to an
arbitrary reference frame is given in the Appendix. The transformation
is needed to implement the elementary amplitudes in nuclear
calculations. 

\section{Eta photoproduction on the nucleon}
\label{free}

Nucleon resonance excitation is the dominant reaction process in
$(\gamma,\eta)$. In contrast to pions which will excite $\Delta(T=3/2)$
as well as $N^*(T=1/2)$ resonances, the $\eta$ meson will only appear in the
decay of  $N^*$ resonances with $T=1/2$. In the low-energy region this is
dominantly the $S_{11}(1535)$ state that decays  45-55\% into $\eta N$,
the only nucleon resonance with such a strong branching ratio in the $\eta$
channel. This result is even more surprising as a nearby resonance of similar
structure, the $S_{11}(1650)$ has a branching ratio of only 1.5\%.
The other resonances that play roles in the low energy region
are the $P_{11}(1440)$ and $D_{13}(1520)$.

Most attempts to describe eta photoproduction on the nucleon\cite{Hic73,Muk91}
have involved Breit--Wigner functions for the resonances and either
phenomenology or an effective Lagrangian approach to model the background. 
These models which contain a large number of free parameters were 
then adjusted to
reproduce the few available data. In a very different approach,
Ref.~\cite{Ben91} derived a dynamical model which employs
$\pi N \rightarrow \pi N,\pi N \rightarrow \pi\pi N$ and
$\pi^- p \rightarrow \eta n$ to fix the hadronic vertex as well as the
propagators and the $\gamma N \rightarrow \pi N$ to construct the
electromagnetic vertex. This calculation represents a prediction rather
than a fit to the $\gamma N \rightarrow \eta N$ reaction.
The model was later extended to include the 
background from $s, u$--channel nucleon Born terms and $\rho,\omega$
exchange in the $t$--channel~\cite{Tia94}. 
Due to the fact that $\eta$ is an isoscalar meson, two coupling schemes 
are possible at the $\eta NN$ vertex: pseudovector(PV) and pseudoscalar(PS).
The latter one is not ruled out by LET as in the case of $(\gamma,\pi)$.
Since the resonance sector is fixed in this approach and 
also the vector meson couplings can be obtained from independent sources,
one can use this model to determine the nature of
$\eta NN$ coupling: both the coupling scheme and the coupling constant.

In contrast to the $\pi N$-interaction, little is known about the $\eta
N$-interaction and, consequently, about the $\eta NN$ vertex.
The uncertainty regarding the structure of the $\eta NN$ vertex
extends to the magnitude of the coupling constant.
This coupling constant $g_{\eta NN}^2/4\pi$ varies between $0$ and $7$
with the large couplings arising from fits of one boson exchange
potentials~\cite{Bro90}.
>From SU(3) flavor symmetry, all coupling constants between the meson octet
and the baryon octet are determined by one free parameter $\alpha$, giving
\begin{equation}
\frac{g_{\eta NN}^2}{4 \pi} = \frac{1}{3}(3-4\alpha)^2\;
\frac{g_{\pi NN}^2}{4 \pi}\;,
\end{equation}
resulting in values for the coupling constant
 between $0.8$ and $1.9$ for commonly used values of
$\alpha$ between $0.6 - 0.65$, depending on the F and D strengths chosen
as the two types of SU(3) octet meson-baryon couplings.
Other determinations of the $\eta NN$
coupling employ reactions involving the eta, such as
$\pi^- p \rightarrow \eta n$, and range from 0.6 - 1.7\cite{Pen87}.
Smaller values are supported by $NN$ forward dispersion
relations\cite{Grein80} with
$g_{\eta NN}^2/{4 \pi}+g_{\eta' NN}^2/{4 \pi} \leq 1.0$.
There is some rather indirect evidence that also favors a small value for
$g_{\eta NN}$.
In Ref.\cite{Pie93}, Piekarewicz calculated the $\pi$--$\eta$
mixing amplitude in the hadronic model where the mixing was generated by
$\bar N N$ loops and thus driven by the proton--neutron mass difference.
To be in agreement with results from chiral perturbation theory the $\eta
NN$ coupling had to be constrained to the range 0.32 -- 0.53.  In a very
different approach, Hatsuda\cite{Hat90} evaluated the proton matrix
element of the flavor singlet axial current in the large $N_C$ chiral
dynamics with an effective Lagrangian that included the $U_A$(1)
anomaly.  In this framework, the EMC data on the polarized proton
structure function (which have been used to determine the "strangeness
content" of the proton) can be related to the $\eta 'NN$ and the $\eta
NN$ coupling constants.  Again, his analysis prefers small values for
both coupling constants.
Nevertheless, from the above
discussion it seems clear that the $\eta NN$ coupling constant is much
smaller than the corresponding $\pi NN$ value of around 14.

\begin{figure}
\centerline{\psfig{file=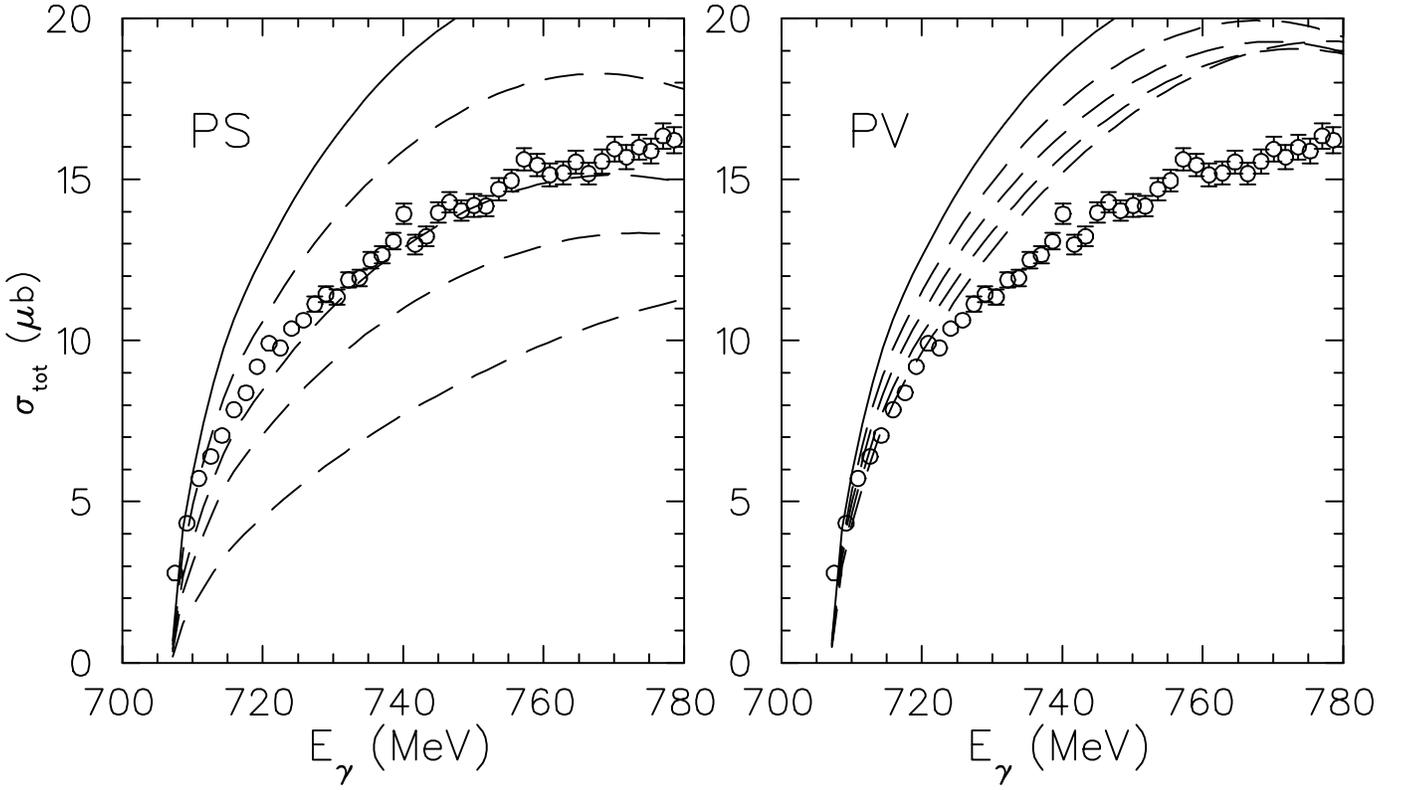}}
\vspace{1cm}
\caption{Total cross section for the process
$(\gamma,\eta)$ on the proton calculated with PS and PV Born terms.
The full curve contains no Born terms, while the dashed lines are (from
the top down) obtained with $g\protect_{\eta NN}^2/4\pi$=0.1, 0.5, 1.0,
and 3.0 for PS-coupling, and
$g\protect_{\eta NN}^2/4\pi$=1.0, 3.0, 6.0 and 10.0 for PV-coupling,
respectively. The experimental data are from Mainz \protect\cite{Kru95}.}
\label{freetot}
\end{figure}

In Fig.~\ref{freetot}, we show the
sensitivity of the total cross section close to threshold when varying
the coupling constant from 0 to 3 for the PS and from 0 to 10 for
the PV form.  There is
a large variation of more than a factor of 2 at 750 MeV for the PS
case while changing the value with PV structure modifies the cross
section only by a relatively small amount.

\begin{figure}
\centerline{\psfig{file=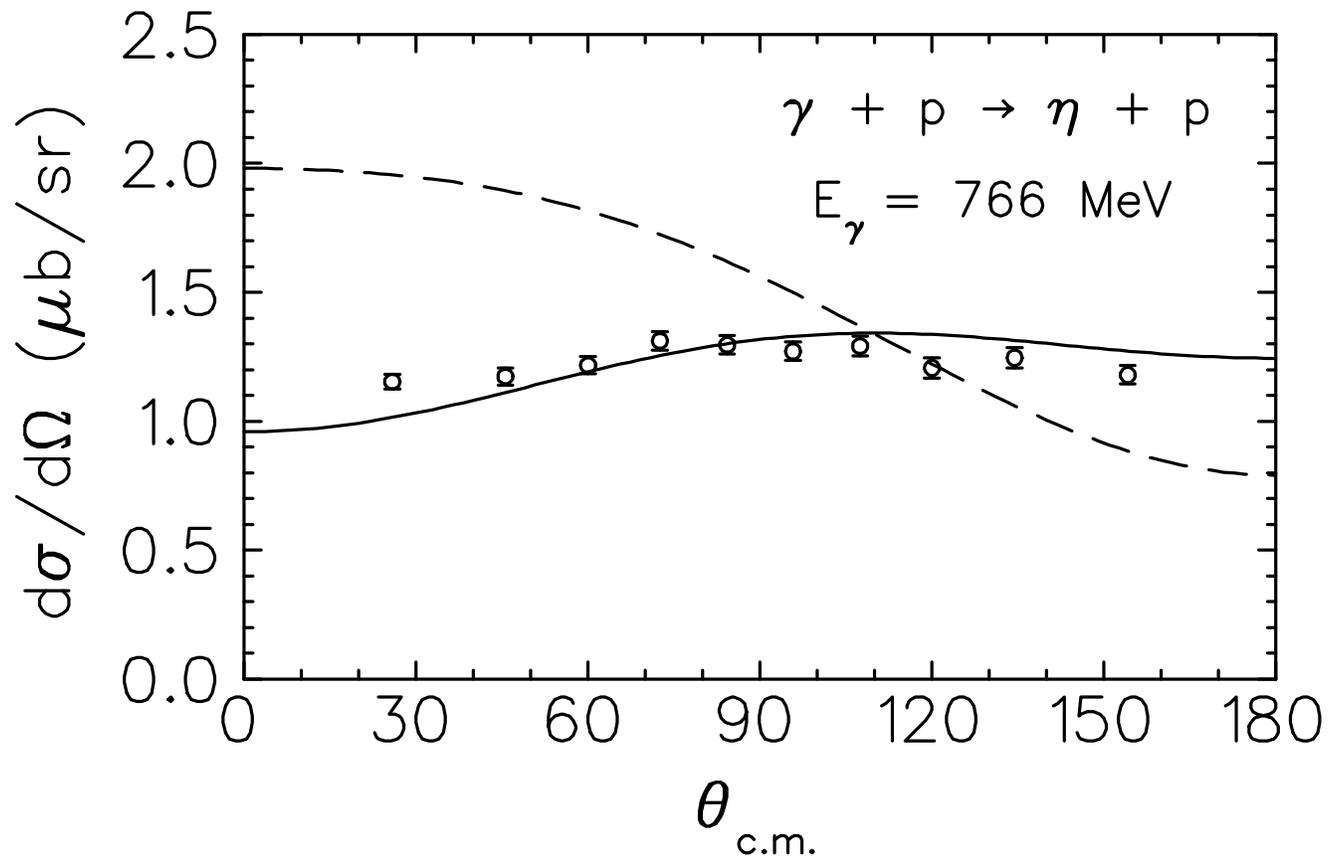}}
\vspace{1cm}
\caption{Differential cross section for eta photoproduction at 766 MeV 
photon energy. The solid and dashed
lines are calculations in PS coupling with coupling constant 0.4 and
in PV coupling with 10, respectively. The data are from Mainz
\protect\cite{Kru95}.}
\label{freedif}
\end{figure}

Here the very precise results of the new Mainz experiment
can clearly distinguish between the different models preferring a
pseudoscalar coupling scheme with
a best-fit coupling constant of $g_{\eta NN}^2/4\pi$=0.4.
This is further supported by comparing with 
the differential cross section as shown in Fig.~\ref{freedif}. 
There is a clear distinction in the forward-backward asymmetry of the 
angular distribution between the PS- and PV-model, and 
PS coupling with $g_{\eta NN}^2/4\pi$=0.4 is strongly supported. 
The variation in the angular distributions is due to the 
$p$-wave multipoles. In particular, the $M_{1-}$ multipole changes 
sign between PS and PV coupling.

\begin{figure}
\centerline{\psfig{file=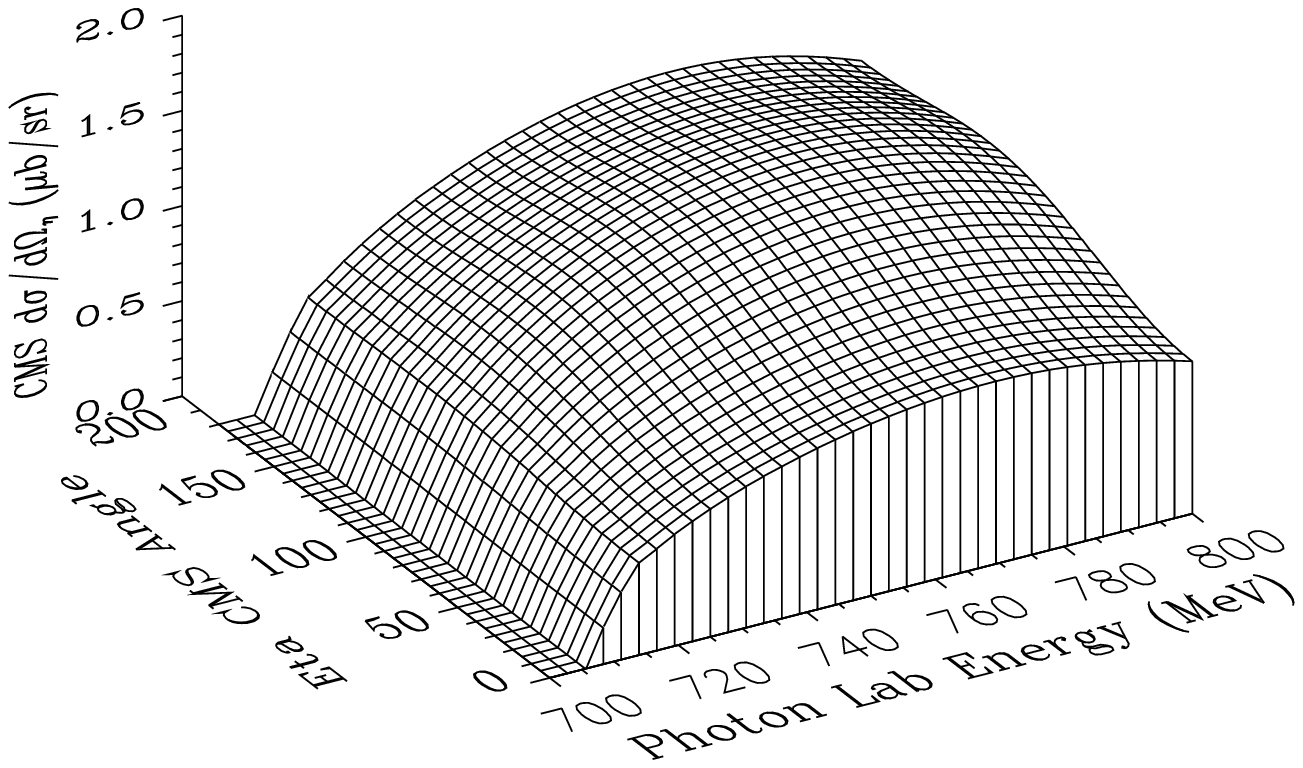}}
\centerline{\psfig{file=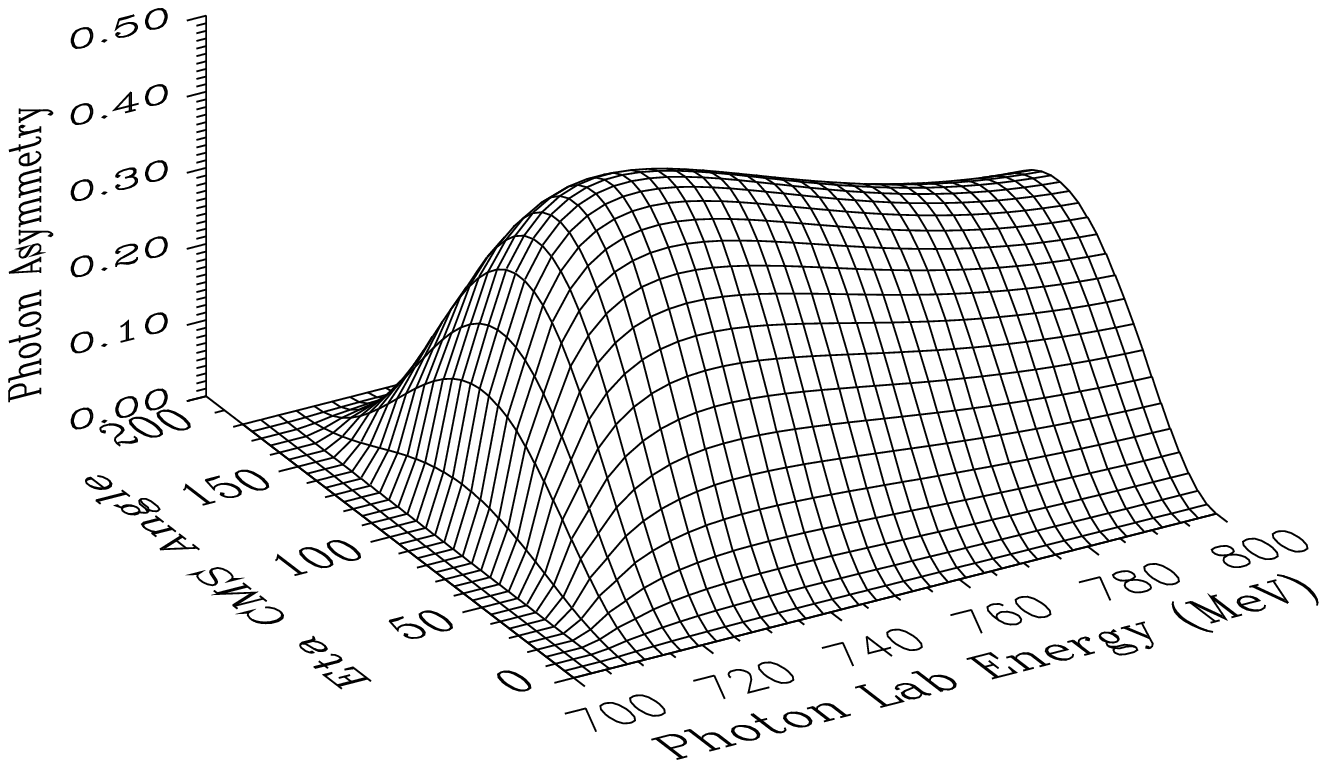}}
\vspace{1cm}
\caption{Complete view of the differential cross section and the
photon asymmetry of eta photoproduction on the proton for PS coupling 
with $g_{\eta NN}^2/4\pi=0.4$.}
\label{f3d}
\end{figure}

Fig.~\ref{f3d} shows the complete view of the differential cross 
section and the photon asymmetry of $\eta$ photoproduction on the 
proton from threshold up to 800 MeV for PS coupling with 
$g_{\eta NN}^2/4\pi$=0.4. 
The resonance dominance near threshold 
is clearly seen in the cross section.
The angular distribution is relatively flat in the cross section but 
peaked in the photon asymmetry.

\section{DWIA model for $\eta$ photoproduction from nuclei}
\label{dwia}

\subsection{Kinematics}

\begin{figure}
\centerline{\psfig{file=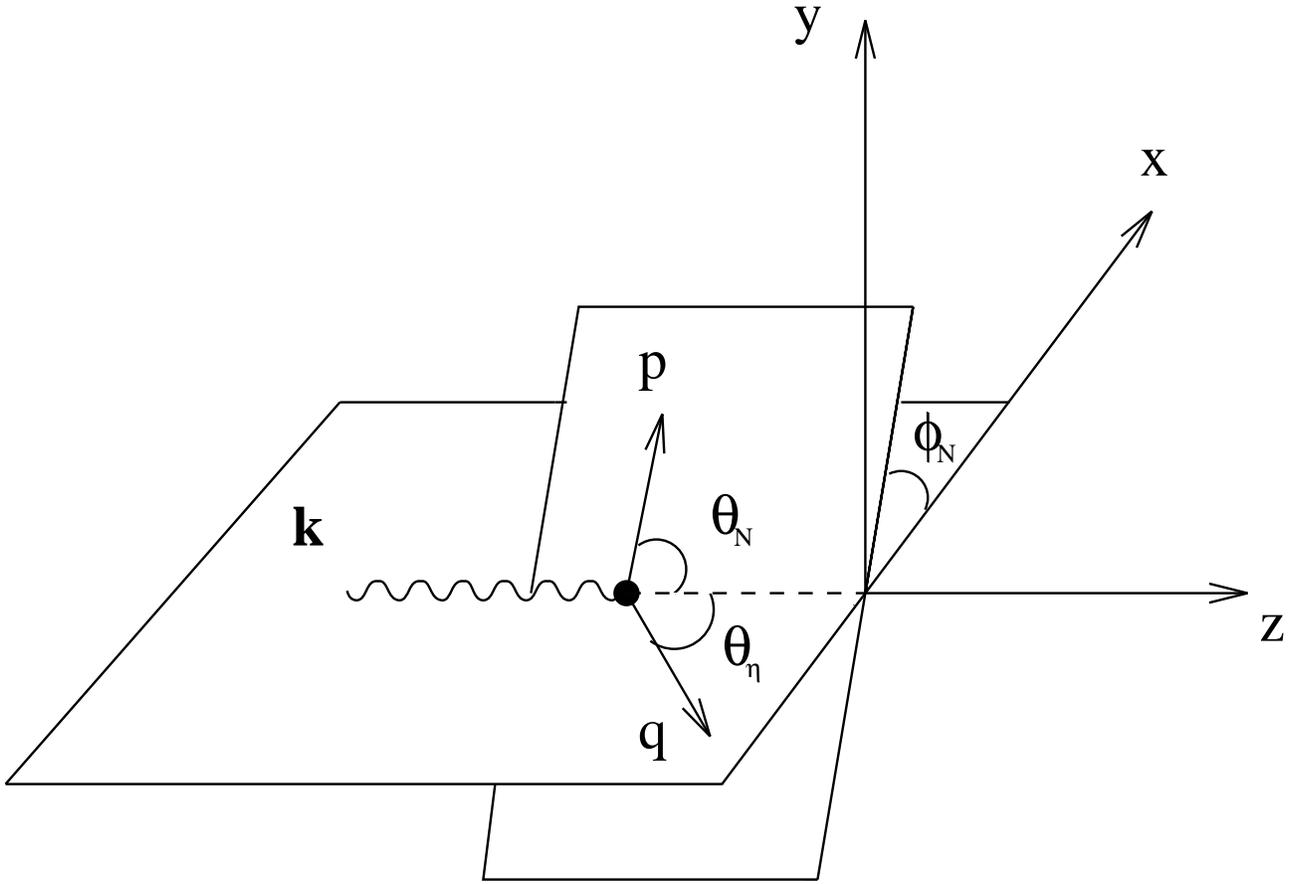}}
\vspace{1cm}
\caption{The coordinate system for the quasifree 
reaction $A(\gamma, \eta N)B$.}
\label{etaxyz}
\end{figure}

In this section we define our coordinate system and discuss some kinematic
aspects of the exclusive reaction $(\gamma,\eta N)$ and the inclusive reaction
$(\gamma,\eta)$ on nuclei.  In the laboratory frame, the four-momenta of the
incoming photon is $k^{\mu}= (E_\gamma,{\bf k})$ while the outgoing eta and
nucleon have four-momenta of $q^{\mu}= (E_\eta,{\bf q})$ and
$p^{\mu}= (E_N, {\bf p})$ respectively.
The target nucleus is at rest with mass $M_i$ and the
recoiling residual nucleus of mass $M_f$ has momentum
\begin{equation}
{\bf Q}={\bf k}-{\bf q}-{\bf p}
\end{equation}
 and kinetic energy $T_Q= \frac{Q^2}{2M_f}$.
The momentum transfer ${\bf Q}$ is sometimes also called the
missing momentum.
Overall energy conservation requires
\begin{equation}
E_\gamma +M_i = E_\eta +E_N +M_f + T_Q.
\label{econ}
\end{equation}
 As shown in Fig.~\ref{etaxyz}, 
the z-axis is defined by the photon direction ${\bf k}$,
and we choose the azimuthal angle of the eta, $\phi_\eta = 0$, by defining
${\bf \hat{y}} = {\bf \hat{k}} \times {\bf \hat{q}}$ and ${\bf \hat{x}} =
 {\bf \hat{y}} \times {\bf \hat{z}}$.

We assume that the reaction takes place on a single bound nucleon with four
-momentum $p{^\mu}{_i} = (E_i, {\bf p}_i)$ and that energy and momentum are
conserved at this vertex (i.e., the impulse approximation).  Thus ${\bf p}_i
=-{\bf Q}$ and $E_i=E_\eta +E_N -E_\gamma$.  If ${\bf Q}$ does not vanish, the
struck nucleon is off its mass shell.  For exclusive reactions this is the
only reasonable off-shell choice since the photon, the eta, the outgoing
nucleon, and the recoiling nucleus are all external lines and must be
on their respective mass shells.  For the inclusive reaction we make the
same choice.

The magnitude of the momentum transfer to the recoiling nucleus has a wide
range, including zero, depending on the directions of the outgoing eta and
nucleon with respect to the incident photon beam.  However, since the
reaction amplitude is proportional to the Fourier transform of the
bound state single particle wavefunction, it becomes quite small for
momentum transfers greater than about 300 MeV/c.  Thus for all but the
lightest nuclei we can safely neglect the nuclear recoil velocity and
generate optical potentials for the outgoing particles in the laboratory
frame.  As noted in \cite{liwb}, reactions of the form $(\gamma, \eta N)$ on
nuclei offer great kinematic flexibility and by appropriate choices one can
investigate the production operator, the bound state wavefunction, or the
final state interaction of the outgoing meson and nucleon.

\subsection{The DWIA cross sections}

In this section we give the DWIA formalism for calculating the
quasifree eta photoproduction  from nuclei.
Following standard procedure, the differential cross section
is given by
\begin{equation}
d\sigma=\frac{1}{2E_\gamma}
\bar{\sum} |{\cal M}_{fi}|^2 (2\pi)^4 \delta^4(P_f+p+q-P_i-k)
\frac{d{\bf q}}{2E_\eta(2\pi)^3} \frac{m_pd{\bf p}}{E_p(2\pi)^3}
\frac{M_f d{\bf Q}}{E_f(2\pi)^3}
\end{equation}
where $\bar{\sum}$ means sums over final spins and
average over initial spins.
The $\delta$-function ensures overall energy-momentum
conservation.
In impulse approximation, the many-body matrix element
can be written as a sum over single particle states
$\alpha=\{nljm\}$
\begin{equation}
{\cal M}_{fi}=\sum_{\alpha}
\langle J_f,m_f|a_\alpha|J_i,m_i \rangle
\langle \eta;N|t_{\gamma \eta}|\alpha;\gamma\rangle
\end{equation}
where $a_\alpha$ is a destruction operator and $t_{\gamma \eta}$
the eta photoproduction on-body operator.
The overlap $S_\alpha=\langle J_f,m_f|a_\alpha|J_i,m_i \rangle^2$
is conventionally called the spectroscopic factor which can
be determined from electron scattering.
After integrating over $d{\bf Q}$ and performing the
sums, one arrives at
the differential cross section
\begin{equation}
\frac{d^3 \sigma}{dE_\eta\,d\Omega_{\eta}\,d\Omega_N}=
\sum_{\alpha,\lambda,m_s}\frac{S_\alpha}{2(2j+1)}
|T(\alpha,\lambda,m_s)|^2
\label{coin}
\end{equation}
where the single particle matrix element
$T=\langle \eta;N|t_{\gamma \eta}|\alpha;\gamma\rangle$ is given by
\begin{equation}
T(\alpha,\lambda,m_s) = \int d^3 r\,
\Psi^{(+)}_{m_s}({\bf r},-{\bf p})\;
\phi^{(+)}_{\eta}({\bf r},-{\bf q})\;
t_{\gamma \eta}(\lambda, {\bf k}, {\bf p}_i ,{\bf q}, {\bf p})\;
\Psi_{\alpha}({\bf r})\;
e^{i{\bf k}\cdot{\bf r}}.  \label{3d}
\end{equation}
In this equation $\lambda$ is the photon polarization, $m_s$
the spin projection of the outgoing nucleon, $\Psi^{(+)}_{m_s}$
and $\phi^{(+)}_{\eta}$ the distorted wavefunctions with outgoing
boundary condition, and $\Psi_{\alpha}$ the bound nucleon
wavefunction.

Another useful observable, the photon asymmetry, is defined by
\begin{equation}
\Sigma=\frac{{d^3 \sigma}_{\perp}-{d^3 \sigma}_{\parallel}}
{{d^3 \sigma}_{\perp}+{d^3 \sigma}_{\parallel}}
\end{equation}
where $\perp$ and $\parallel$ denote the perpendicular and parallel
photon polarizations relative to the production plane (x-z plane).
We have used the short-hand notation
$d^3 \sigma \equiv d^3 \sigma/dE_\eta\,d\Omega_{\eta}d\Omega_N$.

\subsection{The nucleon single particle wavefunction}

As discussed in the section on kinematics, we are primarily interested in
cases of low momentum transfer to the recoiling nucleus.  Thus our choice
of single particle wavefunctions for the bound state is not  critical
as long as the basic size of the orbital is described correctly.  For
convenience we will use harmonic oscillator wavefunctions which have the
advantage that their Fourier transforms are simply obtained.  For each
nucleus under consideration, we adjust the harmonic oscillator
range parameter until the
rms radius of the ground state charge distribution agrees with the
experimentally determined values.

For the continuum nucleon wavefunctions we solve the
Schr\"{o}dinger equation with an optical potential present whenever the
particle is to be detected.  For exclusive reactins, we use the experimentally
determined separation energies for a given orbital in order to fix the
value of the mass of the recoiling nucleus $M_f$.  When we are studying the
inclusive $(\gamma,\eta)$ reaction we allow for all possible final states
by using a plane wave for the proton and taking the mass of the recoiling
nucleus to be equal to the ground state of the A-1 system.  In previous
work on$(e,e')$ reactions from nuclei in the quasielastic
region~\cite{jin1} we found that an assumption similar to this 
described the experimental results very well.

Many optical models for the outgoing nucleon are available.  We use a
non-relativistic reduction of the global optical model in Ref.~\cite{osu}.
This model has the advantage that it fits
nucleon scattering over a wide range of energy and A values, and hence is
very useful for making surveys of a wide range of possible experimental
situations.  Once experimental data is available for the exclusive
reaction, an optical model specific to the nucleus and energy range of the
outgoing nucleon can be substituted.

\subsection{The $\eta$ optical potential}
\label{dw12}

In order to reduce the $\eta$-nucleus many body problem to an 
equivalent potential scattering problem,  we construct a simple 
optical potential. Distorted $\eta$ wavefunctions can be obtained 
by solving the Klein-Gordon equation
\begin{equation}
[-\nabla^2 +\mu^2-\omega^2_\eta]\phi_\eta({\bf r})=
-2\omega_\eta V(r)\phi_\eta({\bf r})
\end{equation}
where $\mu$ is the $\eta$ reduced mass and $\omega_\eta$ its total 
energy. We choose a simple $t\rho$ approximation to construct the 
potential 
\begin{equation}
-2\omega_\eta V(r)=b\rho(r)
\label{pot}
\end{equation}
where $\rho$ is the nuclear local density. This approach is 
justified in the low energy regime since S-waves dominate via
the $S_{11}(1535)$ state and P-wave and D-wave contributions are 
very small. The parameter $b$ is related to the 
$\eta N\rightarrow\eta N$ scattering amplitude by 
\begin{equation}
b=4\pi{p_{\scriptscriptstyle lab} \over p_{\scriptscriptstyle cm}} f
\end{equation}
where $p$ denotes the $\eta N$ two body momentum in the respective frame.
Here we consider two models for the $\eta N$ scattering amplitude.

The first model is from the coupled channel approach 
in Ref.~\cite{Ben91}. We can extract the $\eta N$ t-matrix 
(in the $\eta N$ c.m. frame) by
\begin{equation}
f=\sqrt{2/3}\left[ t_{\scriptscriptstyle S_{11}}
+t_{\scriptscriptstyle P_{11}} \cos\theta
+2t_{\scriptscriptstyle D_{13}}P_2(\cos\theta)\right]/q^0_\eta,
\end{equation}
where $\theta$ is the c.m. angle and 
\begin{equation}
q^0_\eta=[(W^2-(m+m_\eta)^2)(W^2-(m-m_\eta)^2]^{1/2}/{2W}.
\end{equation}
The partial wave scattering amplitudes are given by
\begin{equation}
t^l_{\scriptscriptstyle \eta N\rightarrow\eta N}(W)=
-{1\over 2}\Gamma^l_\eta(W)D(W),
\end{equation}
where $D(W)$ is the resonance propagator (see Eq. 2 in Ref.~\cite{Tia94})
and $\Gamma^l_\eta$ is related to the $\eta$ self-energy 
(see Eq. 3 in Ref.~\cite{Tia94}) by 
$\Gamma^l_\eta=-2\mbox{Im}\Sigma^l_\eta$.
We call the optical potential based on this amplitude DW1.

The second model is given in Ref.~\cite{Bat95}.
In this analysis the $\pi N\rightarrow \eta N$ and 
$\eta N\rightarrow\eta N$ t-matrices are obtained in a 
unitary, coupled, 
three-channel approach with the third channel $\pi\pi N$ 
being an effective two-body channel which
represents all remaining processes. The $\pi N$ elastic 
phase shifts and the weighted data base of the 
 $\pi N\rightarrow \eta N$ total and differential cross sections 
are chosen as the input for the fitting procedure. 
The resulting $\eta N$ t-matrix describes the data fairly well.
For numerical calculations, one can use the following  scattering length
and effective range expansion%~\cite{note}
 which accurately represents the original dominant $S_{11}$ amplitude:
\begin{equation}
f={a_\eta \over 1-i\,a_\eta p_{\scriptscriptstyle cm} 
+{1\over 2}a_\eta r_\eta p_{\scriptscriptstyle cm}^2}
\label{expand}
\end{equation}
with  
\begin{equation}
a_\eta=(0.876+i\,0.274)\; \mbox{fm},\hspace{2mm}
r_\eta=(-1.682-i\,0.139)\; \mbox{fm}.
\end{equation}
We call the optical potential based on this amplitude DW2.  It has been
pointed out in several references~\cite{Wil93,Bat95} that a fit to
total $\pi^-p\rightarrow\eta n$ data based on the optical theorem
requires the imaginary part of the
$\eta N$ scattering length to be larger than roughly 0.22 fm. This would
be difficult to reconcile with DW1 which has $Im{a_\eta}$=0.16 fm or the
model of Ref.~\cite{Bha85} with $Im{a_\eta}$=0.19 fm. The latter two
models were obtained by using $\pi^-p\rightarrow\eta n$  differential
cross section data, some of those have been critized for experimental
problems~\cite{Bat95}. A new $\pi^-p\rightarrow\eta n$ experiment at
Brookhaven\cite{briscoe94} has recently remeasured both differential and
total cross sections at threshold and should clear up this controversy.

%We have performed the same expansion as in Eq.~(\ref{expand})
%for the amplitude in DW1 and find:
%\begin{equation}
%a_\eta=(0.236+i\,0.118)\; \mbox{fm},\hspace{2mm}
%r_\eta=(-5.71-i\,0.391)\; \mbox{fm}.
%\label{dw2}
%\end{equation}

\begin{figure}
\centerline{\psfig{file=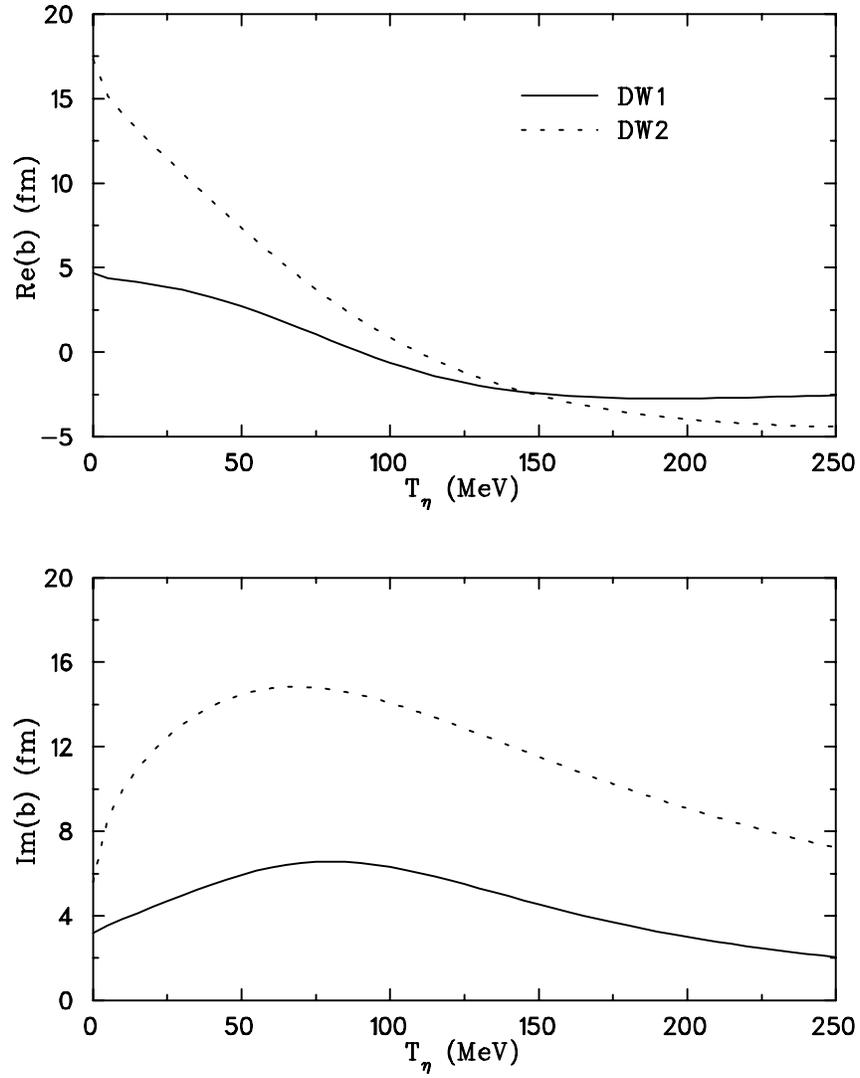}}
\vspace{1cm}
\caption{The energy dependence of the $\eta$ optical potential parameter 
$b$ for two different models of the $\eta N$ t-matrix.
See Sec.~\protect\ref{dw12} for the discussion of DW1 and DW2.}
\label{etapot}
\end{figure}

Fig.~\ref{etapot} shows the energy dependence of the real and 
imaginary parts of the parameter $b$, defined in terms of $f$ in Eq.(11), 
for the two models. We see that the real parts are roughly 
the same for energies larger than 50 MeV, 
but the imaginary parts, which determine absorption, 
are quite different.  As we will see later, it is the imaginary part that most strongly influences inclusive reactions from nuclei.
It is interesting to note that for small eta energies the real part of b is 
positive (more so for DW2 than for DW1) which indicates an 
attractive potential (see Eq.~(\ref{pot})). This has led to the 
suggestion that the $\eta$-nucleus system may form hadronic bound 
states~\cite{liu86,ko89,ch88}.  In the following investigations, unless otherwise noted, we will use the DW1 $\eta$-nucleus optical model.

\section{The exclusive reaction $A(\gamma,\eta N)B$}
\label{exclusive}

The exclusive reaction $A(\gamma,\eta N)B$ where B is in some specific
final state offers a wide range of experimental possibilities.
In order not to be overwhelmed with details, 
we will concentrate on cases where the momentum transfer 
to the recoiling nucleus is allowed to vary freely.
This gives the widest kinematic phase space and 
facilitates discussions on the energy and angular distributions.
Within our model, the reaction occurs on a single nucleon which
is in some specified orbital in the nucleus.  The nucleon can be either a
proton or a neutron and it has some momentum distribution given by the
Fourier transform of its wavefunction.
For simplicity, we consider the coplanar setup
({\em i.e.}, $\phi_\eta=0$, $\phi_N=180^0$) which in general leads to
larger cross sections than out-of-plane setups.

%\subsection{Effects of final-state interactions}

\begin{figure}
\centerline{\psfig{file=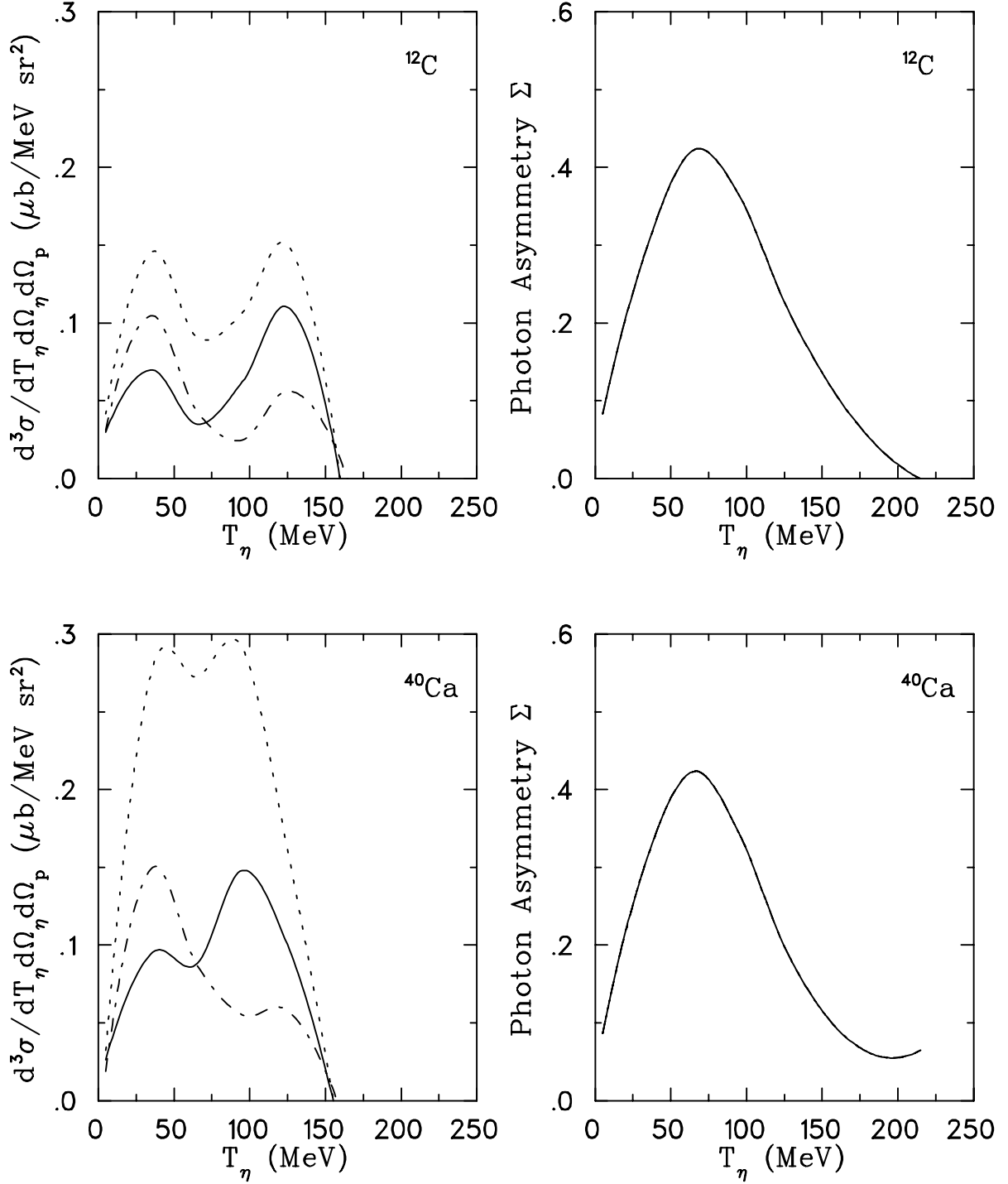}}
\vspace{1cm}
\caption{Effects of distortions in the coincidence cross section 
and the photon asymmetry for $A(\gamma,\eta p),\; p_{3/2}$ at 
$E_\gamma$=750 MeV, $\theta_\eta=20^0$ and $\theta_p=15^0$.
The dashed line is the plane wave calculation, the solid line is with 
$\eta$ distorted by the model DW1 while the proton is kept as plane wave, 
and the dot-dashed line is with $p$ distorted while $\eta$ is 
kept as plane wave.}
\label{gnuc}
\end{figure}

In Fig.~\ref{gnuc}, the effects of final-state $\eta$-nucleus
and p-nucleus interactions from different nuclei are shown
for the exclusive cross section and the photon asymmetry.
The results correspond to knocking out a proton in the $p_{3/2}$ 
orbital of the target nuclei whose momentum distribution is 
largely responsible for the double peaks in the cross sections. 
We see more $\eta$ distortions at lower eta energies than higher eta 
energies, while the opposite is true for proton distortions. 
For fixed photon energy, the proton energy decreases as 
eta energy increases, see Eq.~(\ref{econ}).
As expected, there are larger distortions in $^{40}Ca$ than in $^{12}C$.
It is interesting to note that distortions have no effects on
the photon asymmetry, which is the same result found earlier in investigating exclusive $(\gamma, \pi N)$ reactions from nuclei~\cite{liwb}.
 Furthermore, the photon asymmetry is 
almost independent of the target it is produced from. 
This makes it a sensitive observable for 
studying the production process in the nuclear medium without being 
obscured by distortion effects and overall normalizations. 

\begin{figure}
\centerline{\psfig{file=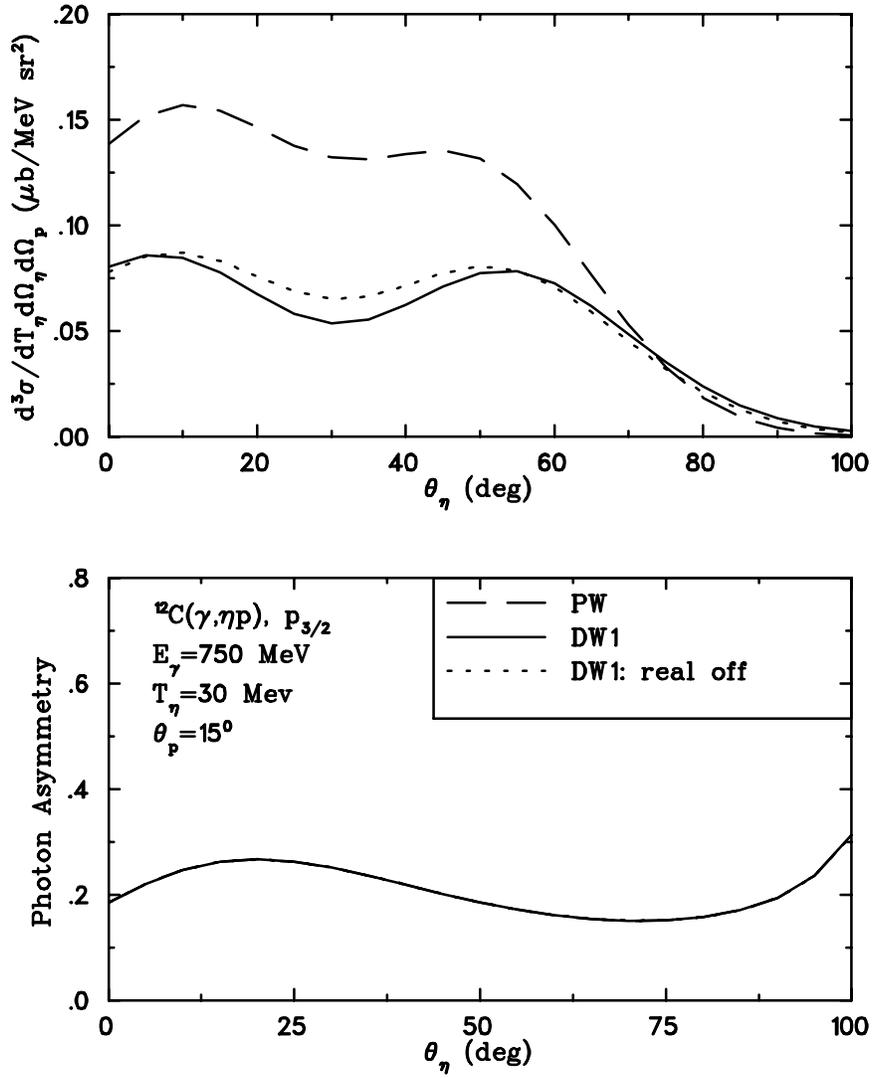}}
\vspace{1cm}
\caption{Effects of the $\eta$-nucleus potential as a function of 
eta angle for plane wave, full DW1, and DW1 with real part turned off.}
\label{ben4}
\end{figure}

In Fig.~\ref{ben4}, the effects of the complete and just the imaginary part of $\eta$-nucleus
final-state interaction are shown as a function of the eta scattering
angle for eta mesons with $30 MeV$ kinetic energy.  Even for this
exclusive reaction, most of the effects on the cross section are due to
the absorptive part of the scattering amplitude. For inclusive
reactions, where the relative angle between the eta and nucleon is not
so important, the effects of the real part of the optical potential will
be even smaller. This insensitivity to the real part of the optical potential 
can be used to determine the imaginary parts of different 
$\eta$-nucleus potentials.
All of these have no effects on the photon asymmetry as
evidenced in Fig.~\ref{ben4} and as discussed above.

%\subsection{Effects of medium modifications}

\begin{figure}
\centerline{\psfig{file=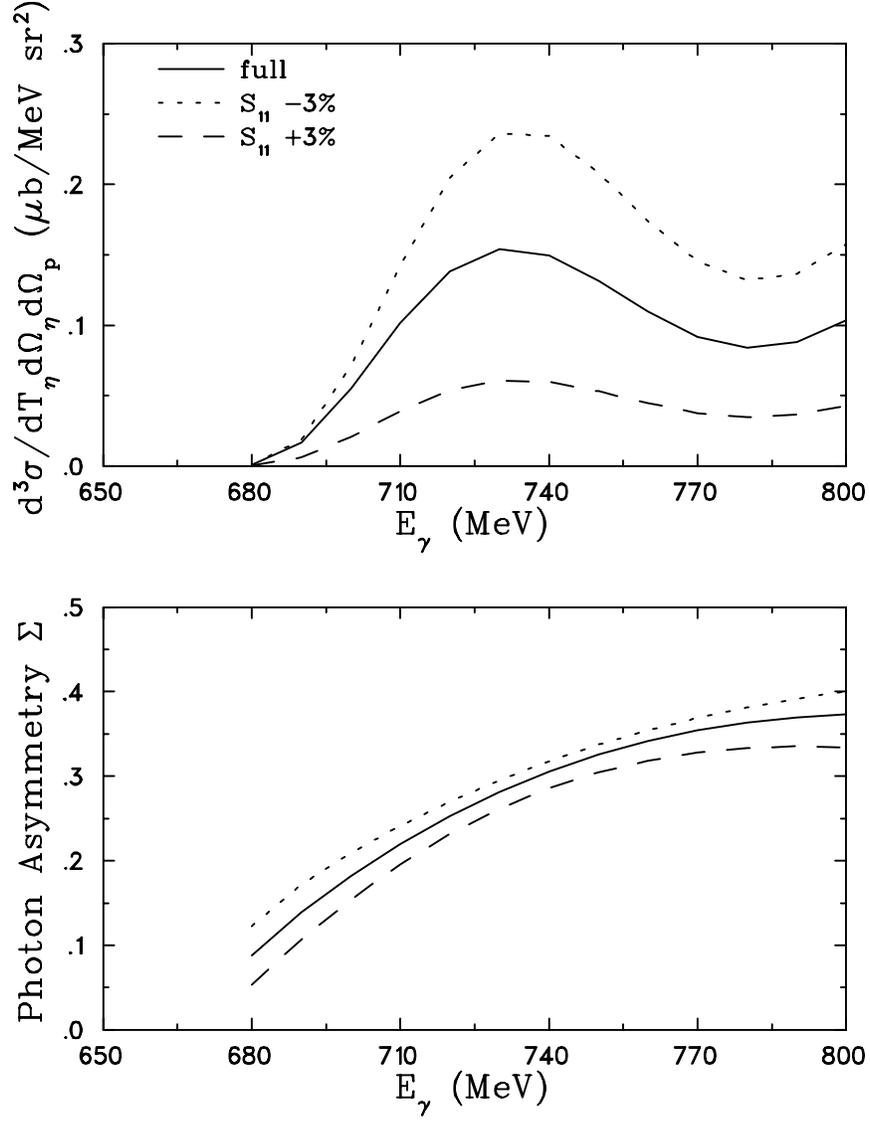}}
\vspace{1cm}
\caption{Effects of medium modifications to the $S_{11}(1535)$ resonance are
shown by changing its bare mass by 3\%. The reaction is
$^{12}C(\gamma,\eta p)^{11}B_{g.s.}$ at $T_\eta$=100 MeV,
$\theta_\eta=20^0$ and $\theta_p=20^0$.}
\label{gs11}
\end{figure}

In Fig.~\ref{gs11} we show the sensitivity to possible 
medium modifications in the cross section and photon asymmetry
for $^{12}C(\gamma,\eta p)^{11}B_{g.s.}$ resulting from creating an
eta from a $p_{3/2}$ proton orbital.
The results are presented as a function of photon energy at fixed 
eta energy and fixed eta and proton angles. 
Under this situation, the proton energy increases and the 
momentum transfer decreases as photon energy increases.
Since we are interested in the relative changes, the results were 
calculated using plane waves.
The solid line is the standard calculation, 
while the dashed and dotted line have the mass of the $
S_{11}(1535)$ increased and decreased by 3\%, respectively. 
We see large effects in the cross section and some
effects in the photon asymmetry.

\begin{figure}
\centerline{\psfig{file=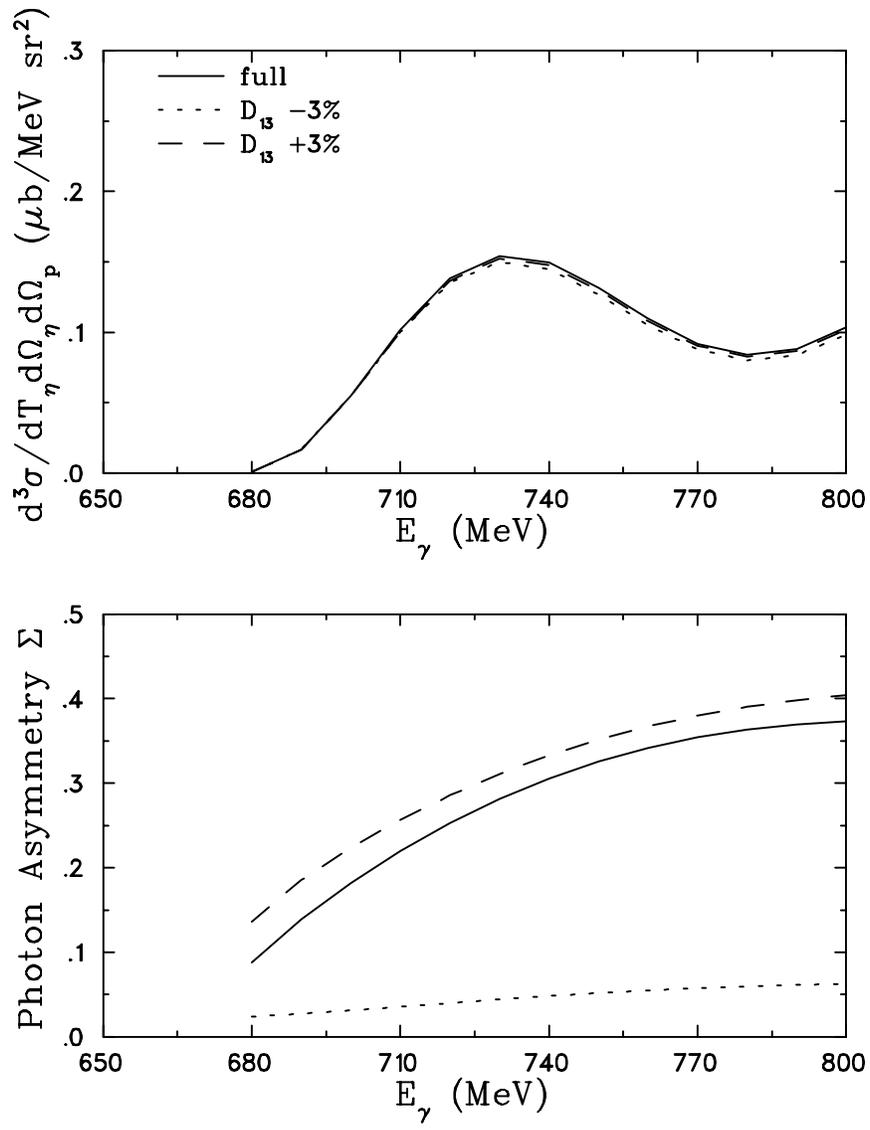}}
\vspace{1cm}
\caption{Same as in Fig.~\protect\ref{gs11}, except for $D_{13}(1520)$.}
\label{gd13}
\end{figure}

Fig.~\ref{gd13} shows the effects of varying the
mass of the $D_{13}$ resonance.
We see large effects in the photon asymmetry while almost no 
effects in the cross section. 
In particular, the photon asymmetry shows a large sensitivity 
to a decrease in the mass of the $D_{13}$. For the bare process
this sensitivity of the photon asymmetry to the $D_{13}$ resonance was
first reported in Ref.~\cite{Tia94}.  We suspect some interplay
between neighboring resonances to be involved.  On the other hand, we found no
sensitivity to varying the mass of the $P_{11}$ resonance.
Clearly, the measurement of the photon asymmetry for the basic reaction and
from nuclei would be very valuable in looking for possible medium modifications. 

\section{The inclusive reaction $A(\gamma,\eta)X$}
\label{inclusive}

Recently, new and good quality data have been obtained at Mainz 
for the inclusive reaction on nuclear targets 
$^{12}C$, $^{40}Ca$~\cite{Roeb95}. In this 
section we will compare our model with the data.

In the inclusive reaction the final nucleon is not observed, 
thus leading to all possible final nuclear states. As mentioned earlier, 
we fix the kinematics for the inclusive reaction by assuming the final 
residual A-1 nucleus is in its ground state, and we use plane waves for the outgoing unobserved nucleon.  In our quasifree model of the reaction, the inclusive 
reaction cross section is obtained by integrating out 
the nucleon solid angle $d\Omega_N$ in Eq.~(\ref{coin}) 
and summing over all nucleon states in  the target
\begin{equation}
{d^2\sigma \over dT_\eta d\Omega_\eta}=
\sum_{i=1}^A\left(\int
\frac{d^3 \sigma}{dE_\eta\,d\Omega_{\eta}\,d\Omega_N}
d\Omega_N\right)_i.
\label{int}
\end{equation}
Note here that the eta production amplitude is different on 
the proton and the neutron. 
In performing the sum over neutron states, we use the ratio 
$\sigma_n/\sigma_p=2/3$ extracted from the Mainz data for
eta photoproduction on the deuteron~\cite{Kru95a}. 

%\subsection{Integrated cross section $d^2\sigma/dT_\eta d\Omega_\eta$}

The kinematics in this case can be solved by 
specifying  the photon energy $E_\gamma$, 
the eta kinetic energy $T_\eta$ and the eta angle $\theta_\eta$.
%Fig.~\ref{g3d} shows a typical 3-dimensional distribution of the  
%cross section at $E_\gamma$=750 MeV. The results correspond to 
%producing an eta from a $p_{3/2}$ proton orbital. 
We examined the distribution of cross sections as a function of 
eta angle and energy at fixed photon energy, and found that most of the
cross sections are concentrated in the forward directions
($\theta_\eta<30^0$),
unlike pion photoproduction. 
It is due to the relatively large mass of the eta as 
compared to the pion. At larger angles, cross sections 
are quickly reduced because of the large momentum transfer involved.
Cross sections as large as around 0.2 $\mu b/MeV sr^2$ can result
at eta energies around 100 MeV. 
Also, etas with energies higher than a certain value (depending on 
the target) are kinematically forbidden.

\begin{figure}
\centerline{\psfig{file=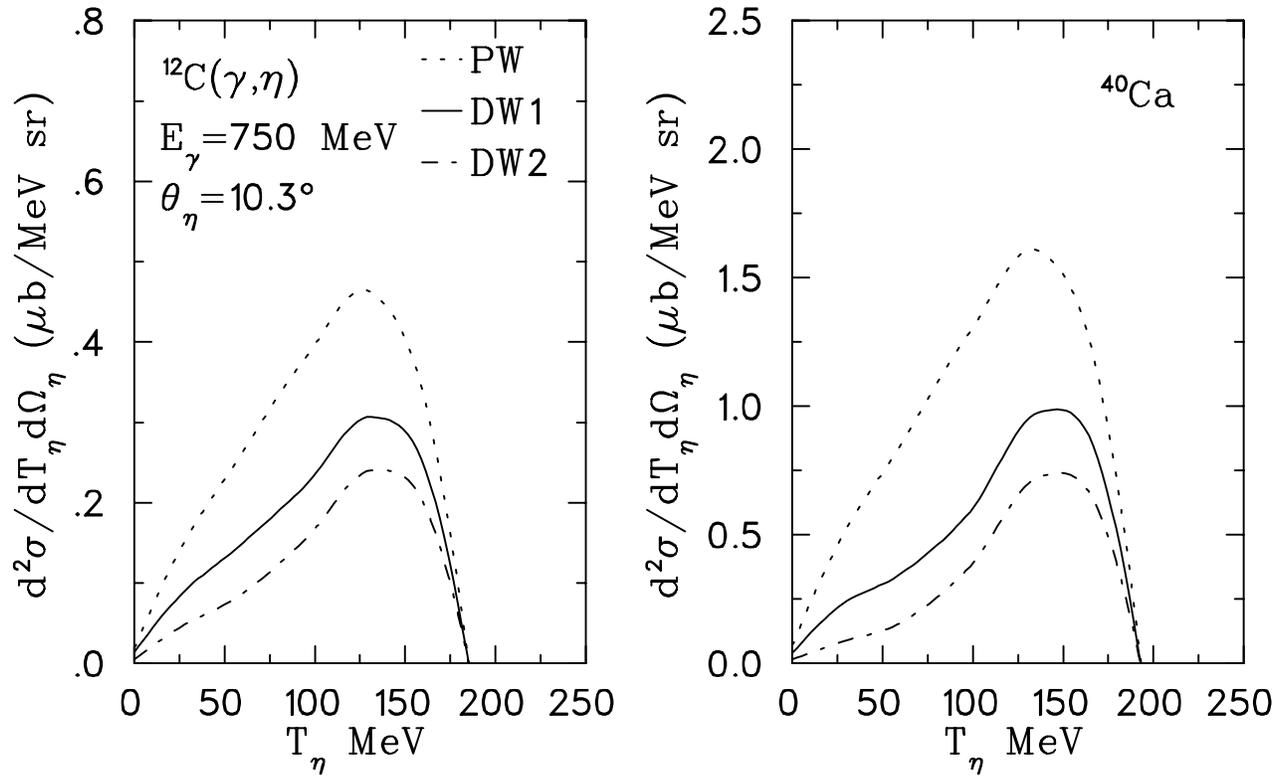,angle=90}}
\vspace{1cm}
\caption{Eta energy dependence of the inclusive cross section 
$d^2\sigma / dT_\eta d\Omega_\eta$
on $^{12}C$ and $^{40}Ca$ calculated in the plane wave (PW) and the 
two different $\eta$-nucleus potentials DW1 and DW2.}
\label{gtth750t}
\end{figure}

\begin{figure}
\centerline{\psfig{file=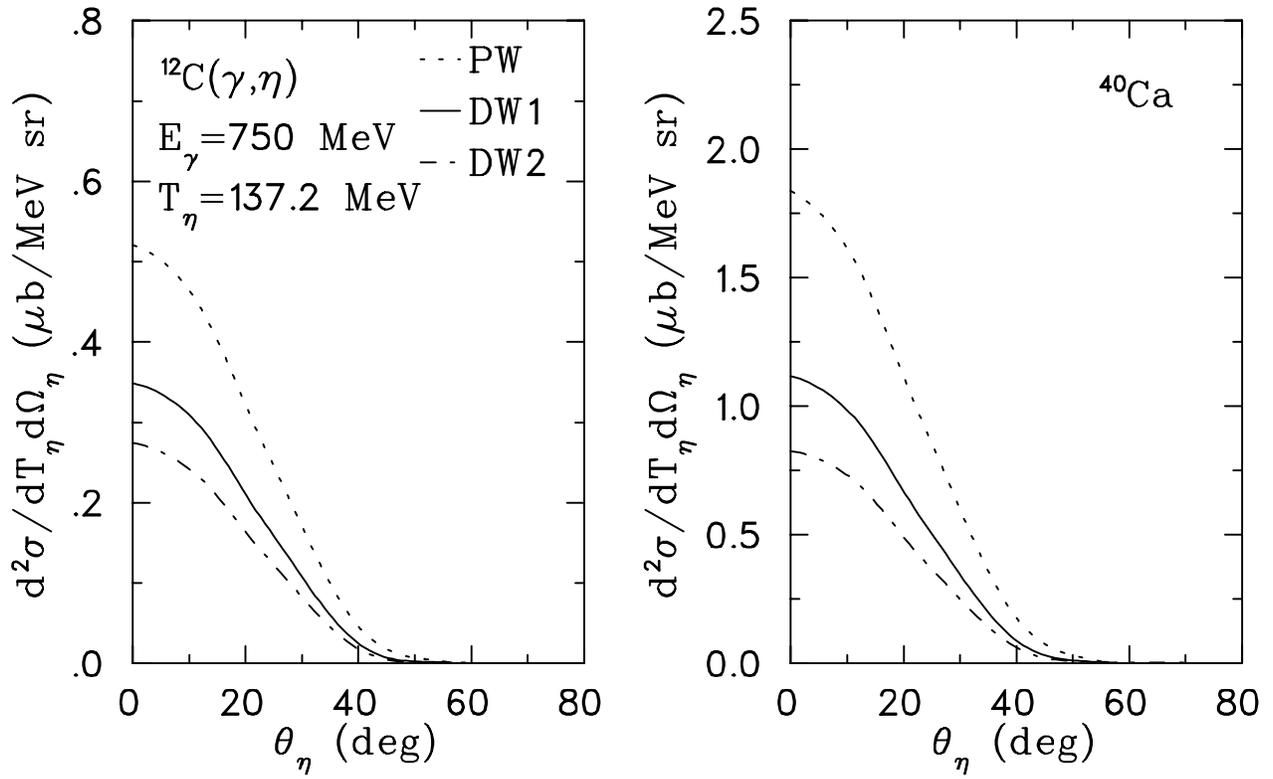,angle=90}}
\vspace{1cm}
\caption{Same as in Fig.~\protect\ref{gtth750t}, but for the 
eta angular dependence.}
\label{gtth750th}
\end{figure}

In Figs.~\ref{gtth750t} and~\ref{gtth750th} we investigate the effect of the two different final state interactions on  two different inclusive cross sections.  In Fig.~\ref{gtth750t} we show the effects of the final-state 
$\eta$-nucleus interaction as a function of eta energy at
fixed photon energy and eta angle, while in Fig.~\ref{gtth750th}  as a function of eta angle at
fixed photon energy and eta energy.  The two different final state interactions investigated are based on models DW1 and DW2 discussed in Section III.D.   The primary result in both cases is a reduction of the cross section due to the absorptive part of t
he optical potential.  As noted earlier, the optical potential based on DW2 is considerably more absorptive than the one based on DW1, which is reflected in these figures.   

%\subsection{Integrated cross section $d\sigma/dT_\eta$}

\begin{figure}
\centerline{\psfig{file=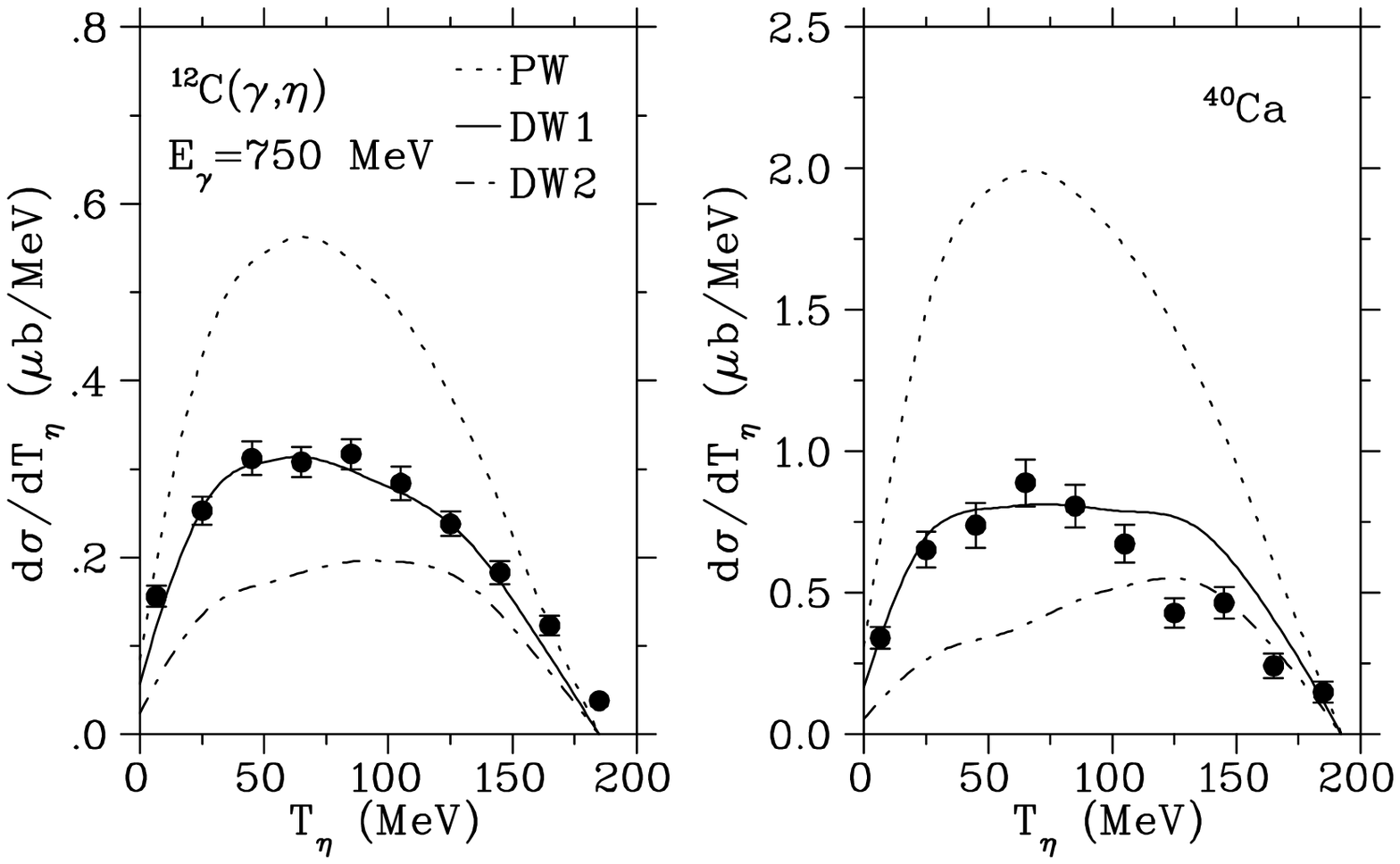,angle=90}}
\vspace{1cm}
\caption{Comparison of our calculations with the Mainz 
data~\protect\cite{Roeb95} 
for the inclusive cross section $d\sigma/dT_\eta $ 
on $^{12}C$ and $^{40}Ca$.}
\label{gt750}
\end{figure}

The integrated cross section $d\sigma/dT_\eta$ is obtained after a further integration over 
the eta solid angle $d\Omega_\eta$ in Eq.~(\ref{int}). 
The kinematics are solved by specifying two variables: 
$E_\gamma$ and $T_\eta$.
Fig.~\ref{gt750} shows the comparison of our calculations with 
the Mainz data on $^{12}C$ and $^{40}Ca$ as a function of the 
eta energy at fixed photon energy.
The plane wave calculation clearly over-predicts the data by about a factor of 2, while the 
distorted wave calculation using DW1 agrees with the data reasonably 
well, and the distorted wave calculation using DW2 underestimates 
the data by about 20 to 30\%. The shape of the energy distribution is
reproduced very well for $^{12}C$ and reasonably well for $^{40}Ca$.  

%\subsection{Integrated cross section $d\sigma/d\theta_\eta$}

\begin{figure}
\centerline{\psfig{file=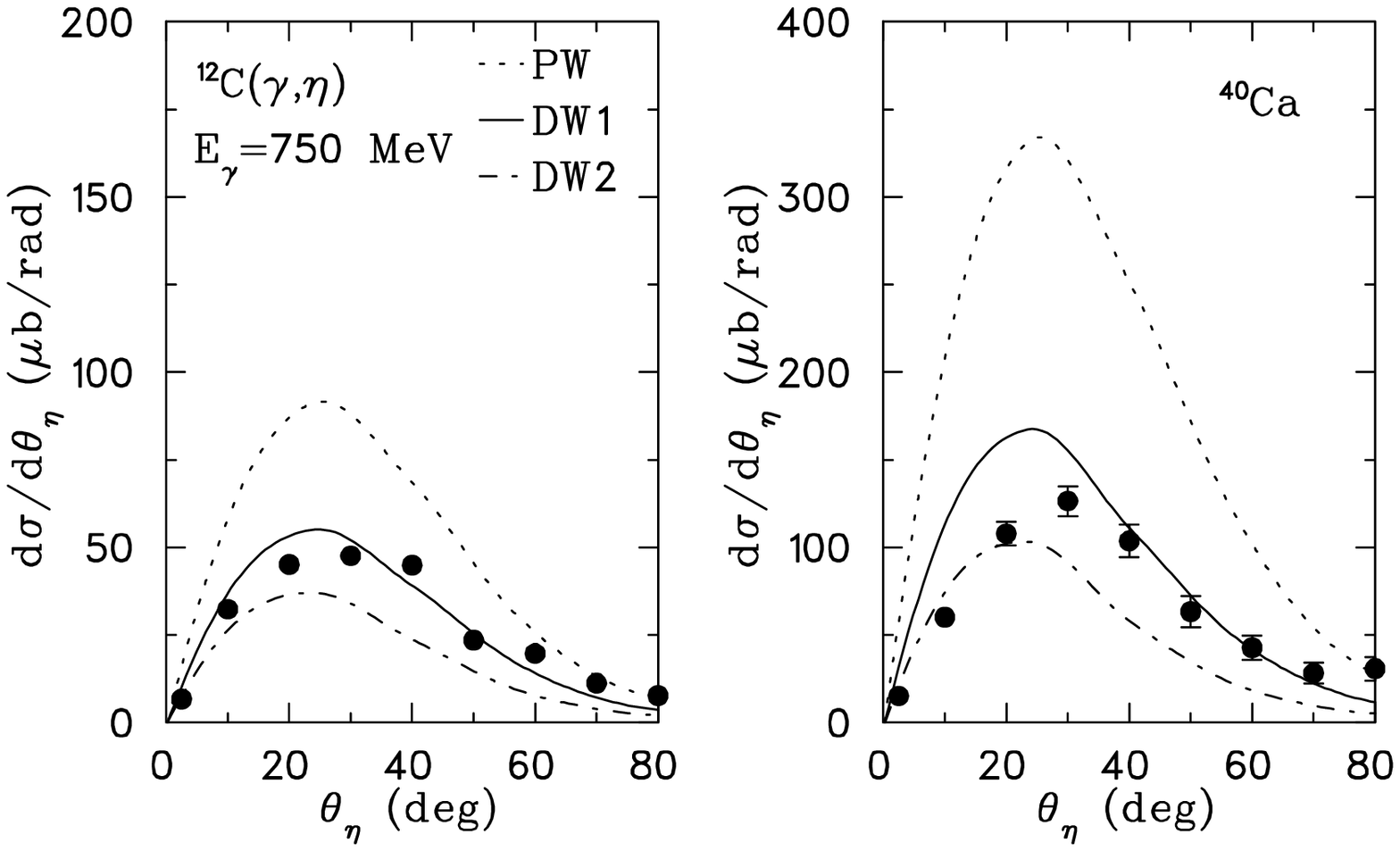,angle=90}}
\vspace{1cm}
\caption{Comparison of our calculations with the Mainz 
data~\protect\cite{Roeb95} for
the inclusive cross section $d\sigma/ d\theta_\eta $ 
on $^{12}C$ and $^{40}Ca$.}
\label{gth750}
\end{figure}

The integrated cross section $d\sigma/d\theta_\eta$is obtained by integrating over the eta energy 
$dT_\eta$ in Eq.~(\ref{int}) and multiplying by the angular factor 
$2\pi\sin\theta$.
The kinematics are solved by specifying two variables: 
$E_\gamma$ and $\theta_\eta$.
Fig.~\ref{gth750} shows the angular dependence at fixed 
photon energy for the comparison of our calculations with 
the Mainz data on $^{12}C$ and $^{40}Ca$.  While the agreement
with $^{12}C$ is also very good here, we predict a somewhat different shape for 
$^{40}Ca$ than is measured.  At lower scattering angles, the DW2  optical model furnishes a better description of the data, while at larger angles, the DW1  optical model is much better.  It could be that the real parts of the optical potentials are playing some role here.

%\subsection{Total cross section}

\begin{figure}
\centerline{\psfig{file=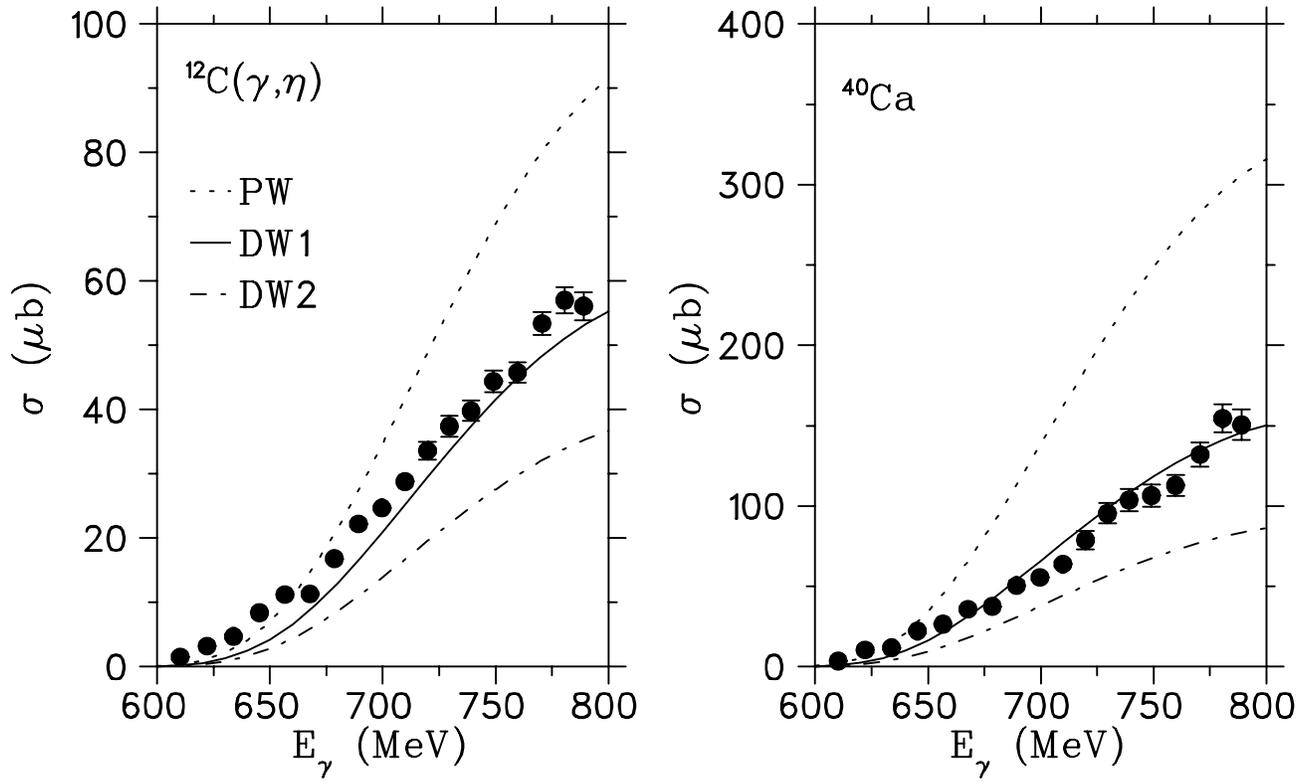,angle=90}}
\vspace{1cm}
\caption{Comparison of our calculations with the Mainz data for
the total cross section on $^{12}C$ and $^{40}Ca$.}
\label{gtot}
\end{figure}

To obtain the total cross section in our model for the quasifree 
eta photoproduction from nuclei, 
we need to perform extensive integrations
which are 4-fold altogether (the $d\phi_\eta$ integration gives 
$2\pi$ due to the azimuthal symmetry after $d\Omega_N$ is performed).
These integrations are done numerically using Gauss' method.
The cross sections are concentrated in a limited region of phase space,
outside of which they are suppressed due to large momentum transfers.
We adjusted the integration limits, the number of integration points, 
and the number of partial waves, until convergence in the total cross 
section is achieved. We arranged the program in such a way that 
the differential cross sections discussed above 
$d^2\sigma/dT_\eta d\Omega_\eta$, $d\sigma/dT_\eta$, 
$d\sigma/d\theta_\eta$, are saved as intermediate results in the 
same run for the total cross section.
We calculated at three photon energies $E_\gamma$=700 MeV, 750 MeV, 
780 MeV and interpolated to other energies since the total cross section
is expected to be a smooth function of $E_\gamma$.
The results are shown in Fig.~\ref{gtot} along with 
experimental data from Mainz.
We see that the model does a good job in reproducing the data.
The distortion (absorption) of etas in the final state is clearly 
needed to achieve this agreement. Again, the absorption present in DW1 is
in better agreement with the data than the absorption in DW2.
The absorption present in DW2, calculated in our model, appears too
strong.

\section{Conclusion}
\label{sum}

We have developed a DWIA model for quasifree eta photoproduction 
on nuclei $A(\gamma,\eta N)B$.
The three key ingredients of the model are  single particle
bound state wave
functions;
the elementary eta photoproduction amplitudes; and
the final state interactions. All of these ingredients enter the 
calculation in a physically transparent way so that one can use 
different models for each ingredient from independent studies. 
Furthermore by using the kinematic flexibility present in the reaction, 
one can emphasize one or more of the ingredients.

In particular, with the exclusive reaction,  kinematics can be chosen so 
that the momentum transfer to the recoiling A-1 nucleus is quite small and 
hence the calculated cross sections are quite insensitive to the bound state
wavefunction.  We found that the exclusive reaction cross sections and,
in particular, the photon asymmetry are very sensitive
to possible medium modifications of the intermediate nucleon resonances.
We recommend that exclusive cross sections be measured, and once
polarized photons are avaiable in this energy range, that the photon asymmetry be 
measured.  This reaction seems to be the cleanest one available for determining
whether or not intermediate resonant states are different in the nuclear medium 
than in free space.

While exclusive cross section data is not yet available, there has been
a series of inclusive cross section measurements from Mainz.  We have compared
our model to the recent inclusive data on $^{12}C$ and  $^{40}Ca$ from Mainz
and we find good agreement in our model
 when we use an $\eta$-nucleus optical potential based on the work
of Bennhold and Tanabe~\cite{Ben91}.  A second optical potential developed in
Ref.~\cite{Bat95} appears to have too much absorption.
We conclude that $\eta$ photoproduction from nuclei provides a good tool
to extract properties of the $\eta A$ interaction.

\acknowledgements
We thank B. Krusche and M. R\"{o}big-Landau for helpful discussions and
for making their data available to us before publication, and A. Svarc and M.
Batinic for providing a convenient expansion of their $\eta N$ amplitude.
F.X.L. was supported by the National Sciences and
Engineering Research Council of Canada;
L.E.W. by U.S. DOE under Grant No. DE-FG-02-87-ER40370,
L.T. by Deutsche Forschungsgemeinschaft (SFB201);
and C.B. by U.S. DOE under Grant No. DE-FG-02-95-ER40907.
We also thank support from the NATO Collaborative Research Grant
and from the Ohio Supercomputing Center for time on the Cray Y-MP.

\newpage
\appendix
\section{Transformation of CGLN amplitudes to an arbitrary reference frame}

Consider a general process of meson photoproduction on the nucleon in
which the final baryon mass $m_2$
can be different from the initial mass $m_1$.
Energy-momentum conservation gives $k+p_1=q+p_2$.
The kinematics can be described by the
Lorentz-invariant Mandelstam variables
$s=(k+p_1)^2$, $t=(k-q)^2$, $u=(k-p_2)^2$ with
the constraint $s+t+u=m_1^2+m_2^2+m_q^2$,
or by the invariant mass $W=\sqrt{s}$ and the scattering angle $\theta$.

In the center-of-mass (CMS) frame we denote the four-momenta as
$k=(E_\gamma,{\bf k})$, $p_1=(E_1,-{\bf k})$,
$q=(E_q,{\bf q})$, $p_2=(E_2,-{\bf q})$.
These CMS variables can be expressed in terms of W:
$E_1=(W^2+m_1^2)/(2W)$, $E_2=(W^2+m_2^2-m_q^2)/(2W)$, etc.
The cross section in the CMS frame can be written as
\begin{equation}
d\sigma/d\Omega=(k/q)|{\cal T}_{fi}|^2
\label{norm}
\end{equation}
where we have used
$k=|{\bf k}|$ and $q=|{\bf q}|$ to denote
the initial and final 3-momentum in the system, respectively.

The transition matrix element can be expressed as
\begin{equation}
{\cal T}_{fi}=\langle\chi_2|F|\chi_1\rangle
\label{cg}
\end{equation}
where $\chi_1$, $\chi_2$ are Pauli spinors and
\begin{equation}
F=i\sigma\cdot\epsilon\;F_1 + \sigma\cdot \hat{\bf q}\;
\sigma\cdot(\hat{\bf k}\times \epsilon)\;F_2
+i\sigma\cdot \hat{\bf k}\;\sigma\cdot\epsilon\;F_3
+i\sigma\cdot\hat{\bf q}\;\hat{\bf q}\cdot\epsilon\;F_4
\end{equation}
The F's are the CGLN amplitudes~\cite{cgln}.
Note that there is a factor $m_1m_2/(4\pi W)$ absorbed
in the F's under the normalization of Eq.~(\ref{norm}).

On the other hand, the transition matrix element can be expressed in the invariant form
\begin{equation}
{\cal T}_{fi}=\sum_{j=1}^{4} A_j\bar{u}(2)M_j u(1)
\label{inv}
\end{equation}
where $u(2)$ and $u(1)$ are Dirac spinors, A's the invariant
amplitudes and M's the gauge and Lorentz invariant matrices defined by
\begin{eqnarray}
M_1 &=& -\gamma_5 \epsilon\!\!/ k\!\!/   \\
M_2 &=& 2\gamma_5 ( \epsilon\cdot p_1\, k\cdot p_2 -
\epsilon\cdot p_2\, k\cdot p_1)  \\
M_3 &=& \gamma_5 (\epsilon\!\!/ k\cdot p_1 -k\!\!/ \epsilon\cdot p_1) \\
M_4 &=& \gamma_5 (\epsilon\!\!/ k\cdot p_2 -k\!\!/ \epsilon\cdot p_2)
\end{eqnarray}

The F's are only defined in the CMS frame, but the A's are valid in
 an arbitrary frame.
One can find the relations between the F's and the A's  by
first expanding Eq.~(\ref{inv}) in Pauli space, then expressing the result
in the CMS frame and putting it in the form of Eq.~(\ref{cg}).
After some algebra, we find
\begin{eqnarray}
F_1 &=& \frac{k}{4\pi}\left(\frac{E_2+m_2}{2W}\right)
\left[-cA_1+\frac{W+m_1}{2}cA_3-\frac{u-m_2^2}{2(W-m_1)}cA_4\right]  \\
F_2 &=& \frac{k}{4\pi}\left(\frac{E_2-m_2}{2W}\right)
\left[cA_1+\frac{W-m_1}{2}cA_3-\frac{u-m_2^2}{2(W-m_1)}cA_4\right]  \\
F_3 &=& \frac{kq}{4\pi}\left(\frac{E_2+m_2}{2W}\right)
\left[(W-m_1)cA_2-cA_4\right] \nonumber \\
F_4 &=& \frac{kq}{4\pi}\left(\frac{E_2-m_2}{2W}\right)
\left[-(W+m_1)cA_2-cA_4\right]
\end{eqnarray}
where $c=\sqrt{(E_1+m_1)(E_2+m_2)/(4m_1m_2)}$.

Inverting these equations, we obtain
\begin{eqnarray}
cA_1 &=& \frac{4\pi}{k\sqrt{2W}}
\left[-\frac{W-m_1}{\sqrt{E_2+m_2}}F_1+\frac{W+m_1}{\sqrt{E_2-m_2}}F_2\right]
\\
cA_2 &=& \frac{4\pi}{kq\sqrt{2W}}
\left[\frac{F_3}{\sqrt{E_2+m_2}}-\frac{F_4}{\sqrt{E_2-m_2}}\right] \\
cA_3 &=& \frac{4\pi}{k\sqrt{2W}}
\left[\frac{2F_1}{\sqrt{E_2+m_2}}+\frac{2F_2}{\sqrt{E_2-m_2}}
-\frac{u-m_2^2}{q^2}\left(\frac{\sqrt{E_2-m_2}}{W-m_1}F_3
+\frac{\sqrt{E_2+m_2}}{W+m_1}F_4\right) \right] \\
cA_4 &=& \frac{4\pi}{kq\sqrt{2W}}
\left[\frac{-F_3}{(W+m_1)\sqrt{E_2+m_2}}
-\frac{F_4}{(W-m_1)\sqrt{E_2-m_2}} \right].
\end{eqnarray}
Substituting the A's back into Eq.~(\ref{inv}), we obtain a
transition operator which is valid in any reference frame and 
which is suitable for use in nuclear calculations.
Eq.~(\ref{inv}) can be further expanded into Pauli space and cast into
the form of a spin non-flip term plus a spin flip term
\begin{equation}
{\cal T}_{fi}=L+i\,{\mbox{\boldmath $\sigma$}} \cdot  {\bf K}.
\end{equation}

\newpage

\end{document}